\def\beq{\begin{equation}}
\def\eeq{\end{equation}}
\def\blg{\begin{align}}
\def\elg{\end{align}}
\def\beg{\begin{gather}}
\def\eeg{\end{gather}}
\def\bea{\begin{eqnarray}}
\def\eea{\end{eqnarray}}
\def\bed{\begin{displaymath}}
\def\eed{\end{displaymath}}
\def\bef{\begin{figure} \begin{center}}
\def\eef{\end{center} \end{figure}}

\def\no{\noindent}

\def\1{\'{\i}}

\documentclass[preprint,pre,aps]{revtex4-1}
\usepackage{amssymb,amsmath,graphics,graphicx,bm}

\begin{document}

\title{Phase synchronization of coupled bursting neurons and the generalized Kuramoto model}
\author{F. A. S. Ferrari $^1$, R. L. Viana $^1$ \footnote{Corresponding author. e-mail: viana@fisica.ufpr.br}, S. R. Lopes $^1$, and R. Stoop $^2$}
\affiliation{1. Departament of Physics, Federal University of Paran\'a, 81531-990 Curitiba, Paran\'a, Brazil \\
2. Institute of Neuroinformatics, University of Z\"urich and Eidgenössische Technische Hochschule Z\"urich, 8057 Z\"urich, Switzerland}

\date{\today}

\begin{abstract}
Bursting neurons fire rapid sequences of action potential spikes followed by a quiescent period. The basic dynamical mechanism of bursting is the slow currents that modulate a fast spiking activity caused by rapid ionic currents. Minimal models of bursting neurons must include both effects. We considered one of these models and its relation with a generalized Kuramoto model, thanks to the definition of a geometrical phase for bursting and a corresponding frequency. We considered neuronal networks with different connection topologies and investigated the transition from a non-synchronized to a partially phase-synchronized state as the coupling strength is varied. The numerically determined critical coupling strength value for this transition to occur is compared with theoretical results valid for the generalized Kuramoto model.  
\end{abstract}

\maketitle

\section{Introduction}

Neurons are known to exhibit a plethora of dynamical behaviors, represented by the generation of action potential patterns. One of such patterns is bursting, defined by the repeated firing of action potentials followed by quiescent periods. Hence the dynamics of bursting neurons has two timescales: a fast scale related to spiking and a slow scale of bursting itself. These timescales are related to different biophysical mechanisms occuring at the level of neuron membrane: there are fast ionic currents (chiefly $Na^+$ and $K^+$) responsible for spiking activity and slower $Ca^{++}$ currents that modulate this activity.

Most neurons exhibit bursting behavior if conveniently stimulated. For example, in the neocortical layer $5$ pyramidal neurons, when stimulated with DC current pulses, fire an initial burst of spikes followed by shorter bursts \cite{connors, ruedi01}. In layers $2$, $3$, and $4$ chattering neurons fire high-frequency bursts of $3-5$ spikes with a short interburst period \cite{gray, ruedi02}. Cortical interneurons have been found to exhibit bursting as a response to DC pulses \cite{markram}. Pyramidal neurons in the $CA1$ region of hippocampus produce high-frequency bursts after current injection \cite{su}. Thalamocortical neurons and reticular thalamic nucleus inhibitory neurons exhibit bursting as well \cite{ramcharan}. Purkinje cells in cerebellum can burst when their synaptic input is blocked \cite{womack}. Bursting is also an important feature of sensory systems, because bursts can increase the reliability of synaptic transmission \cite{krahe}. In some systems, bursts improve the signal-to-noise ratio of sensory responses and might be involved in the detection of specific stimulus features \cite{metzner}.

Due to both synaptic coupling and common inputs among neurons there are many types of synchronization, which can be generally regarded as the presence of a consistent temporal relationship between their activity patterns \cite{1,2,3}. A strong form of the latter relationship is complete synchronization, where neurons spikes at the same time, i.e. a precise temporal coincidence of events. A weaker relationship is bursting synchronization, in which only the beginning of bursting is required to occur at the same time, even though the repeated spiking may not occur synchronously. 

There has been observed bursting synchronization in cell cultures of cortical neurons, where uncorrelated firing appeared withing the first three days and transformed progressively into synchronized bursting within a week \cite{kamioka}. Large-scale bursting synchonization in the $7 - 14 Hz$ range has been found in the thalamus during slow-wave sleep, partially originated in the thalamus and gated by modulatory input from the brainstem \cite{steriade1}. Various areas of the basal ganglia have been found to exhibit bursting synchronization related to Parkinson's disease and resting tremor \cite{bevan}. 

There exists sound neurophysiological evidence that hypokinetic motor symptoms of Parkinson’s disease such as slowness and rigidity of voluntary movements are closely related to synchronized bursting in the $10-30 Hz$ range \cite{uhlhaas,brown,hutchison,park}. The connection between bursting synchronization and pathological conditions like Parkinson's disease, essential tremor and epilepsy has led to the proposal of many control strategies aiming to suppress or mitigate bursting synchronization \cite{hammond}. 

One of such strategies is deep-brain stimulation (DBS), which consists of the application of an external high-frequency ($> 100 Hz$) electrical signal by depth electrodes implanted in target areas of the brain like the thalamic ventralis intermedius nucleus or the subthalamic nucleus \cite{benabid}. The effect of DBS would be similar to that produced by tissue lesioning and has proved to be effective in suppression of the activity of the pacemaker-like cluster of synchronously firing neurons, and achieving a suppression of the peripheral tremor \cite{blond}. There is strong clinical evidence that DBS is a highly effective technique for treatment of patients with Parkinson's disease \cite{rodriguez,albanese}.

In spite of these results, DBS is yet far from being completely understood. Many results in this field have been obtained from empirical observations made during stereotaxic neurosurgery, but further progress can be obtained with proper mathematical modelling of DBS \cite{tass,pfister,hauptmann}. The effects of DBS in networks of bursting neurons have been investigated when DBS is implemented through an harmonic external current \cite{batista1} and a delayed feedback signal \cite{batista2}. 

On modelling the response of a neuronal network to an external perturbation like DBS it is of paramount importance to keep the model simple enough such that large-scale simulations (using a large number of neurons) can be performed in a reasonable computer time. In such reductionist point of view a minimal model could be one in which we can assign a geometrical phase to the bursting activity. The bursting neuron is thus regarded as a phase oscillator undergoing spontaneous oscillations with a given frequency. Thus bursting synchronization becomes a special case of phase synchronization, a phenomenon well understood for coupled oscillators with and without external excitation \cite{synchrobook}. 

A simple model for the dynamics of nonlinearly coupled phase oscillators is the Kuramoto model, which in its original version considers a global (all-to-all) coupling \cite{kuramoto}. It can be generalized by considering an arbitrary coupling architecture (generalized Kuramoto model) \cite{acebron}. The particular interest in such models is that many analytical and numerical results are known for them, specially the global case for which a mean-field theory exists for the transition between a non-synchronized to a (phase-)synchronized behavior \cite{strogatz}. For generalized Kuramoto models it is possible to derive analytical expressions for the critical value of the coupling strength for which the abovementioned transition occurs \cite{restrepo}. Hence such a body of knowledge can be applied to networks of bursting neurons, helping to design strategies of synchronization control and/or suppression like DBS.

The main goal of this paper is to show, using analytical and numerical arguments, that a system of coupled bursting neurons described by Rulkov's model can be reduced to a generalized Kuramoto model. This reduction is valid as long as phase synchronization is concerned, since for frequency synchronization the behaviors can be quite different, though. We consider, in particular, some widely used connection topologies, like random (Erd\"os-Renyi), small-world, and scale-free networks. We show that the analytical results for the critical coupling strength to synchronized behavior, originally derived for the generalized Kuramoto model, can be used to describe the synchronization transition also for networks of bursting neurons. 

As a matter of fact, since bursting activity presents two timescales it can be also approached from the point of view of a relaxation oscillator \cite{wang}. In our work, however, we describe bursting using a single phase. This simplification is justified since phase synchronization of bursting is chiefly related to the slow timescale. In other words, the fast spikes can be nonsynchronized even though the slow dynamics is synchronized.

This paper is organized as follows: in Section II we describe the model we used to describe bursting neurons. Section III considers networks of coupled bursting neurons using different connection topologies. Section IV reviews some results on the generalized Kuramoto model, and Section V includes the comparisons we made between Kuramoto model and the network of bursting neurons. Our Conclusions are left to the final Section. 

\section{Models of bursting neurons}

The choose of a suitable model describing the dynamics of biological neurons is dictated by some requirements. First the model must take into account the kind of dynamics one wishes to describe \cite{miguel}. For example, if all one needs is to describe a spiking neuron, for the sake of neural coding simulations for example, a simple leaky integrate-and-fire (LIF) model would be enough \cite{koch}. However if one needs to describe the interplay between different ionic currents flowing through the neuron membrane, the Hodgkin-Huxley (HH) model would be a natural choice \cite{hodgkin}. On the other hand, the HH model would require far more computational power than the LIF since the former involves four complicated first-order differential equations whereas the latter just one simple equation. 

With bursting neurons this criterion also holds. Given that bursting results from the interplay between fast and slow ionic currents, Hodking-Huxley-type models would need at least one more equation to describe slow $Ca$ modulation \cite{plant,shorten}. A model of thermally sensitive neurons exhibiting bursting has been proposed by Huber and Braun \cite{braun98,braun00,braun01}, which describes spike train patterns experimentally observed in facial cold receptors and hypothalamic neurons of the rat \cite{braun99}, electro-receptors organs of freshwater catfish \cite{schafer95}, and caudal photo-receptor of the crayfish \cite{feudel00}. However, the Huber-Braun model has $5$ differential equations for each neuron, and computational limitations impose restrictions to its use for large networks \cite{nosso}. 

If numerical simulations do not need to take into account the effect of system parameters and only the phenomenological aspects of bursting are relevant, then a good choice is the two-dimensional mapping equations proposed by Rulkov \cite{rulkov01}
\begin{eqnarray}
\label{rulkovx}
x(n+1) & = & \frac{\alpha}{1 + {[x(n)]}^2} + y(n), \\
\label{rulkovy}
y(n+1) & = & y(n) - \sigma x(n) - \beta,
\end{eqnarray}
\noindent where $x$ is the fast and $y$ is the slow dynamical variable, whose values are taken at discrete time $t = n \tau$, with $\tau = 1$ and $n = 0, 1, 2, \ldots$. 

\begin{figure}
\begin{center}
\includegraphics[width=0.6\textwidth,clip]{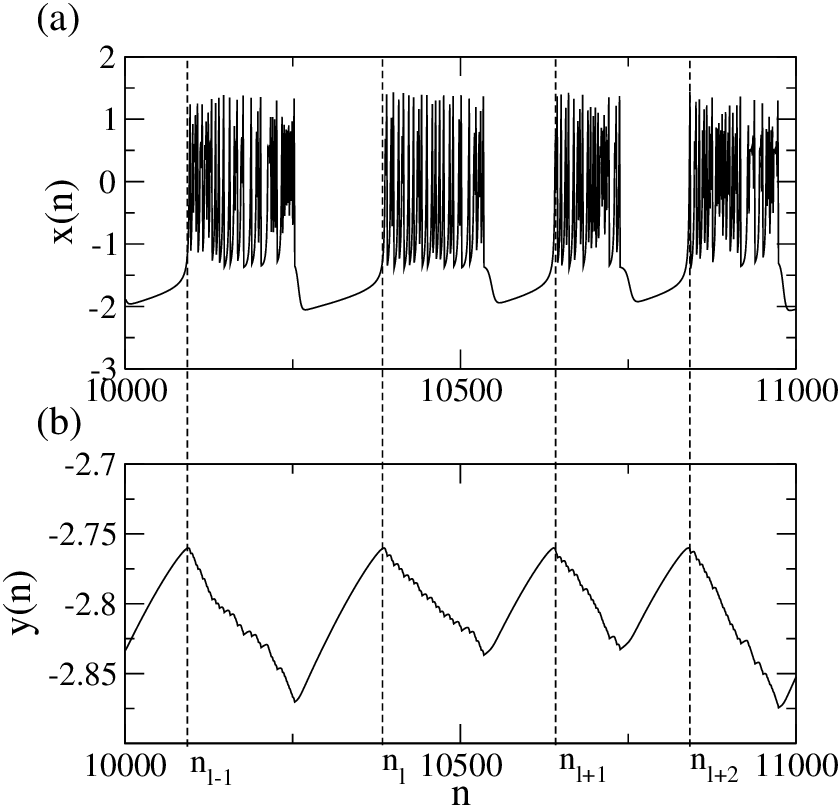}
\end{center}
\caption{\label{rulkovfig} Time evolution of the (a) fast and (b) slow variables in the Rulkov map (\ref{rulkovx})-(\ref{rulkovy}) for $\alpha = 4.1$, $\sigma = \beta = 0.001$.}
\end{figure}

The parameter $\alpha$ affects directly the spiking timescale, its values being chosen in the interval $[4.1, 4.3]$ so as to produce chaotic behavior for the evolution of the fast variable $x_n$, characterized by an irregular sequence of spikes.  The parameters $\sigma$ and $\beta$, on their hand, describe the slow timescale represented by the bursts, and take on small values (namely $0.001$) so as to model the action of an external dc bias current and the synaptic inputs on a given isolated neuron \cite{rulkov04}. 

The Rulkov model was derived using dynamical rather than biophysical hypotheses. We choose the parameter $\alpha$ so as to yield chaotic behavior for the characteristic spiking of the fast variable $x(n)$ [Fig. \ref{rulkovfig}(a)]. The bursting timescale, on the other hand, comes about the influence of the slow variable $y(n)$, which provide a modulation of the spiking activity due to a saddle-node bifurcation [Fig. \ref{rulkovfig}(b)]. On comparing the dynamics of the Rulkov map with similar results of the Huber-Braun model one is led to an approximate correspondence between variables of the Rulkov map and variables with biophysical significance: the discrete time $n$ in the map corresponds to $0.5 ms$ of the continuous time; and $x$ stands for the membrane potential, in such a way that each unit of $x$ in Fig. \ref{rulkovfig}(a) corresponds to {\it circa} $20 mV$. The Rulkov model has been used in several numerical investigation of coupled neuron models \cite{rulkov01,7,8,9}

Due to the chaoticity of the $x$-dynamics of the Rulkov map, the duration of each burst suffers a slight variability. The beginning of each burst can be chosen rather arbitrarily, but it turns out that a useful choice is to consider the local maxima of the variable $y$: let $n_\ell$ be the time at which the $\ell$th burst begins [see Fig. \ref{rulkovfig}(b)]. Hence the duration of this burst is $n_{\ell+1} - n_\ell$. We can define a geometric bursting phase by considering a variable $\varphi$ that increases of $2\pi$ after a bursting event. A linear interpolation gives \cite{ivanchenko}
\beq
\label{phase}
\varphi(n) = 2\pi \ell + 2\pi \frac{n-n_\ell}{n_{\ell+1}-n_{\ell}}, \qquad (n_\ell \le n \le n_{\ell+1}).
\eeq
We can also define a bursting (angular) frequency, which gives the time rate of the phase evolution:
\begin{equation}
\label{frequency}
\omega = \lim_{n\rightarrow\infty} \frac{\varphi(n) - \varphi(0)}{n}.
\end{equation}

\section{Networks of coupled Rulkov neurons}

In the Rulkov model (\ref{rulkovx})-(\ref{rulkovy}) the variable $x(n)$ plays the role of the membrane potential at discrete time $t = n \tau$, where $\tau = 1$. Hence the difference $x(n+1)-x(n)$ can be interpreted as the time derivative of the potential. If the membrane capacitance is scaled to the unity, it amounts to the transmembrane current. Hence the effect of coupling is to inject a synaptic current in the equation for the $x$-variable (\ref{rulkovx}). 

Let us denote by $x_i$ the membrane potential of the $i$th neuron ($i=1, 2, \ldots N$). The equations for a network of coupled neurons are
\begin{eqnarray}
\label{x}
x_i(n+1) & = & \frac{\alpha_i}{1+{[x_i(n)]}^2} + y_i(n) + \varepsilon \sum_{j=1}^{N} A_{ij} 
x_j(n), \\
\label{y}
y_i(n+1) & = & y_i(n) - \sigma x_i(n) - \beta,
\end{eqnarray}
in which $A_{ij}$ is the adjacency matrix, which defines the type of synapse which connects the $i$th and $j$th neurons (respectively post-synaptic and pre-synaptic). 

If the spatial distance between them is taken into account, we can model an electrical synapse by a band-diagonal adjacency matrix which includes only near-neighbors of a given neuron. Chemical synapses are thus represented here by off-band-diagonal elements, since they include the effect of distant neurons. This is a simplified model, though, since it does not take into account the number of open channels in the post-synaptic neuron.  For studies of synaptic-dependent phenomena, like plasticity, memory storage, learning, pattern recognition, etc. this simplified form of the coupling term may not be sufficient for numerical simulations. Moreover, we consider all interactions as bidirectional, in such a way that the adjacency matrix is symmetric ($A_{ji}=A_{ij}$), what is more likely to be the case in electrical than chemical synapses. However these simplifications are acceptable in a model as long as we are interested only in the effect of the network topology on the neuron dynamics. An example is the investigation of the DBS effects related to tremor associated with bursting synchronization in the thalamus \cite{daniela}.

The only parameter in our model that is to characterize the intensity of the synaptic connections is the coupling strength $\varepsilon$. It takes on positive values for excitatory synapses and negative values for inhibitory ones. Since $\varepsilon$ is the same for all neurons in our model we cannot consider here the case in which part of the synapses are excitatory ($ \sim 75 \%$) and part inhibitory. A modification such as this, although simple to implement in principle, would make the model more difficult to compare with minimal models of phase dynamics. The only restriction on the values of $\varepsilon$ is that the coupling term itself cannot be too large so as to drive the neuron dynamics off a bursting state. This parameter must be adjusted by trial-and-error.

Since in neuronal assemblies the neurons are likely to exhibit some diversity, we choose randomly the values of $\alpha$ within the interval $[4.1, 4.3]$ according to a specified probability distribution function (PDF) $\tilde{g}(\alpha)$ \cite{rulkov02}. In spite of this we keep the other parameters with the same values (namely $\sigma=\beta=0.001$) since this do not change our results in an appreciable way. We can compute the bursting phase $\varphi_n^{(i)}$ and the corresponding angular frequencies $\omega^{(i)}$ using (\ref{phase}) and (\ref{frequency}), respectively, for the coupled neurons in the same way as we did for isolated ones. 

When the $\alpha$ parameter is chosen in the interval $[4.1,4.3]$ there is a linear relation between $\alpha$ and the mean burst frequency. For this reason when we choose the $\alpha$ parameter for each neuron according with a given PDF $\tilde{g}(\alpha)$ we generate a PDF for the bursting frequency of uncoupled neurons denoted by $g(\Omega)$ with the same shape as ${\tilde g}(\alpha)$. In this work we consider two types of distributions: (i) a waterbag (or uniform) distribution:
\beq
\label{uniform}
{\tilde g}_W(\alpha) = \left \{ 
\begin{array}{ccc}
   \frac{1}{a-b} & \mbox{for} & a\leq \alpha \leq b \\
   0 & \mbox{otherwise}
\end{array}
\right.,
\eeq
where $a=4.1$ and $b=4.3$; and (ii) a truncated Cauchy distribution:
\beq
\label{cauchy}
{\tilde g}_T(\alpha) = \frac{1}{\gamma} {\left\lbrack 
\tan^{-1} \left(\frac{b-\alpha_0}{\gamma}\right) - \tan^{-1} \left(\frac{a-\alpha_0}{\gamma}\right) \right\rbrack}^{-1} 
{\left\lbrack 1 + {\left(\frac{\alpha-\alpha_0}{\gamma}\right)}^2 \right\rbrack}^{-1},
\eeq
where $a=4.1$, $b=4.3$, $\alpha_0=4.2$ is the position of the distribution peak, and $\gamma=0.1$ is the half-width at half-maximum of the PDF \cite{nadarajah}.

\begin{figure}
\begin{center}
\includegraphics[width=0.7\textwidth,clip]{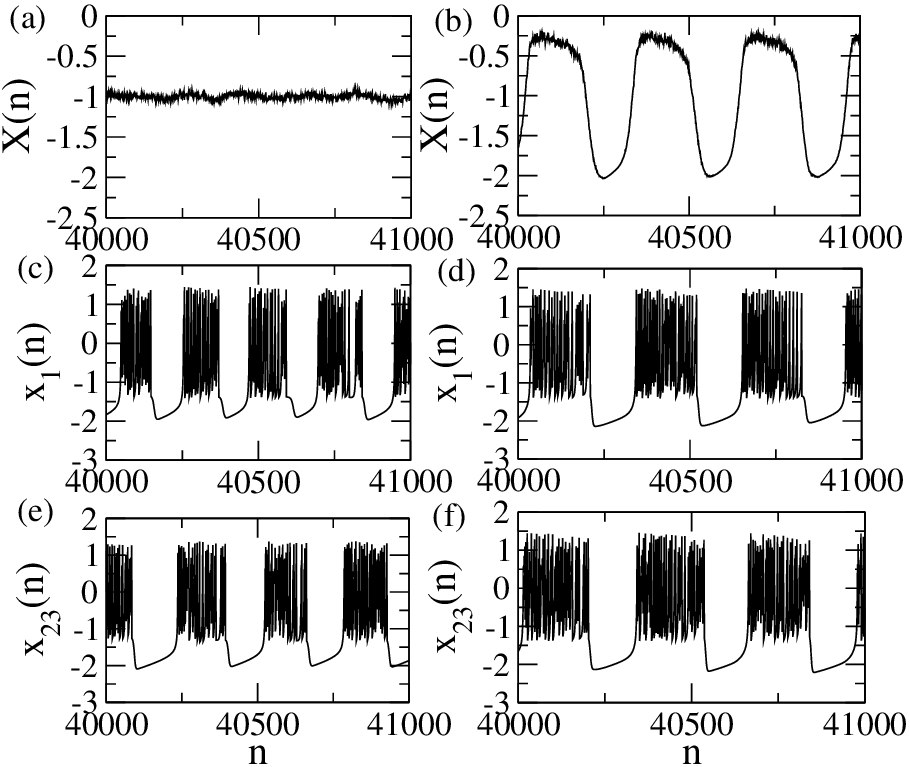}
\end{center}
\caption{\label{bs} Time evolution of (a) mean field and (c,e) fast variables of two selected neurons in a network of uncoupled Rulkov neurons. (b,d,f) stand for coupled neurons.}
\end{figure}

We shall defer the discussion of the nature and properties of the adjacency matrix to the following section. For the moment let us assume that a convenient form of $A_{ij}$ has been given to the system of coupled Rulkov neurons (\ref{x})-(\ref{y}). One distinctive effect of synaptic coupling is bursting synchronization: two or more neurons begin their bursting activity at the same times (up to a given tolerance), regardless of whether or not the ensuing sequence of spikes coincides. A example is provided by Fig. \ref{bs}, where we consider two randomly selected neurons in a network: in the uncoupled case the neurons burst at different and uncorrelated times [Fig. \ref{bs}(c) and (e)], what makes the mean field
\beq
\label{meanfield}
X(n) = \frac{1}{N} \sum_{i=1}^N x_i(n),
\eeq
to have small-amplitude fluctuations [Fig. \ref{bs}(a)]. For large enough values of $\varepsilon$ the coupled system of neurons present bursting synchronization as illustrated by the two selected neurons [Fig. \ref{bs}(d) and (f)]. The mean field in this case undergoes large-amplitude oscillations with the same frequency as of the bursting neurons [Fig. \ref{bs}(b)].

Another quantitative characterization is provided by the order parameter magnitude and its time average given, respectively, by \cite{kuramoto}
\beq
\label{order}
R(n) = \frac{1}{N} \left\vert \sum_{i=1}^{N} e^{i\varphi_i(n)} \right\vert,
\qquad
\overline{R} = \frac{1}{n'} \sum_{n=1}^{n'} R(n),
\eeq
where $n'$ is chosen so as to yield stationary values of $R$. If the neurons burst exactly at the same time their phases coincide and hence the normalized sum in (\ref{order}) gives $R = 1$. On the other hand, if the neuron bursting is so uncorrelated that the times at which they burst are practically random the summation gives nearly zero and $R \approx 0$ for such an extreme case.  In finite networks there is likely to be chance correlations, hence we consider $R = 0.1$ as a threshold for the transition from a non-synchronized to a partially synchronized state. Intermediate cases ($0.1 < R < 1$) thus represent partially synchronized states.

\section{Connection architectures of the neuronal network}

Neurons are connected by axons and dendrites, so that we can regard those neurons as embedded in a three-dimensional lattice. However, due to the high connectivity of the neurons it is necessary to use complex networks to describe the properties of neuronal assemblies. These complex networks can be viewed in two basic levels: a microscopic, neuroanatomic level, and a macroscopic, functional level. Studies in the former level are limited to those few examples in which there is available data on the neuronal connectivity, as the worm {\it C. Elegans} \cite{varshney}. 

In the second (macroscopic) level of description of neural networks, the use of non-invasive techniques as electroencephalography, magnetoencephalography and functional magnetic resonance imaging provides anatomical and functional connectivity patterns between different brain areas \cite{hilgetag,gorka2}. This information provides a way to study the brain cortex, considering the latter as being divided into anatomic and functional areas, linked by axonal fibers. Scannell and coworkers have investigated the anatomical connectivity matrix of the visual cortex for the macaque monkey and the cat \cite{scannell1,scannell2}.

The basic elements of a complex network are nodes and links. A link connects two or more nodes, and these connections can be unidirectional or bidirectional. In the language of neuronal  networks a node can be either a single neuron or a cortical area, depending on the level of characterization. If we consider a microscopic description the links are synapses (electrical or chemical), whereas they are axonal fibers in networks of cortical areas or simply they stand for the functional relationship between the corresponding regions in the cortex.

A complex network can be characterized by various topological and metrical properties \cite{barabasi}. For our purposes we will need only two of them: the average path length $L$ and the average clustering coefficient $C$. The path length between two nodes ${\bf a}$ and ${\bf b}$ is the minimum number of links that must be traversed through the network to travel from ${\bf a}$ to ${\bf b}$. The average path length $L$ is the path length averaged over all pairs of nodes in the network. The average clustering coefficient $C$ of a network is defined as the average fractions of pairs of neighbors of a given node which happen also to be neighbors of each other \cite{watts}. For example, if a node ${\bf a}$ is connected to nodes ${\bf b}$ and ${\bf c}$, and if ${\bf b}$ and ${\bf c}$ are themselves connected, then ${\bf a}$, ${\bf b}$ and ${\bf c}$ form a triangular cluster. The value of $C$ turns to be the fraction of triangular clusters that exist in the network with respect to the total possible number of such clusters.

From the graph-theoretical perspective a complex network can be described by an adjacency matrix $A_{ij}$, whose elements are equal to the unity if the nodes are connected and zero otherwise. In studies of cortical area networks, both in the anatomic and functional levels, each link is sometimes given a specific weight, which can be even time-varying in investigations of plasticity. Since we discard self-interactions, the diagonal elements of the adjacency matrix are zero.

\subsection{Global and random networks}

In a globally coupled network all nodes are connected with all other nodes. Since in this case the contribution of the coupling would increase with the number of nodes, the coupling term is usually divided by the number of nodes. In such a network each node can be regarded as being coupled to the mean field of other nodes. The adjacency matrix of global networks is constant: $A_{ij} = 1$ for all $i,j=1, 2, \ldots N$. There have been numerical investigations of such networks with respect to strategies of DBS \cite{ivanchenko,rosenblum}. The coupling term in (\ref{x}) will be divided by the number of neurons, hence we shall redefine it as 
\beq
\label{xi}
\varepsilon = \frac{\xi}{N}
\eeq

\begin{figure}
\includegraphics[width=1.0\textwidth,clip]{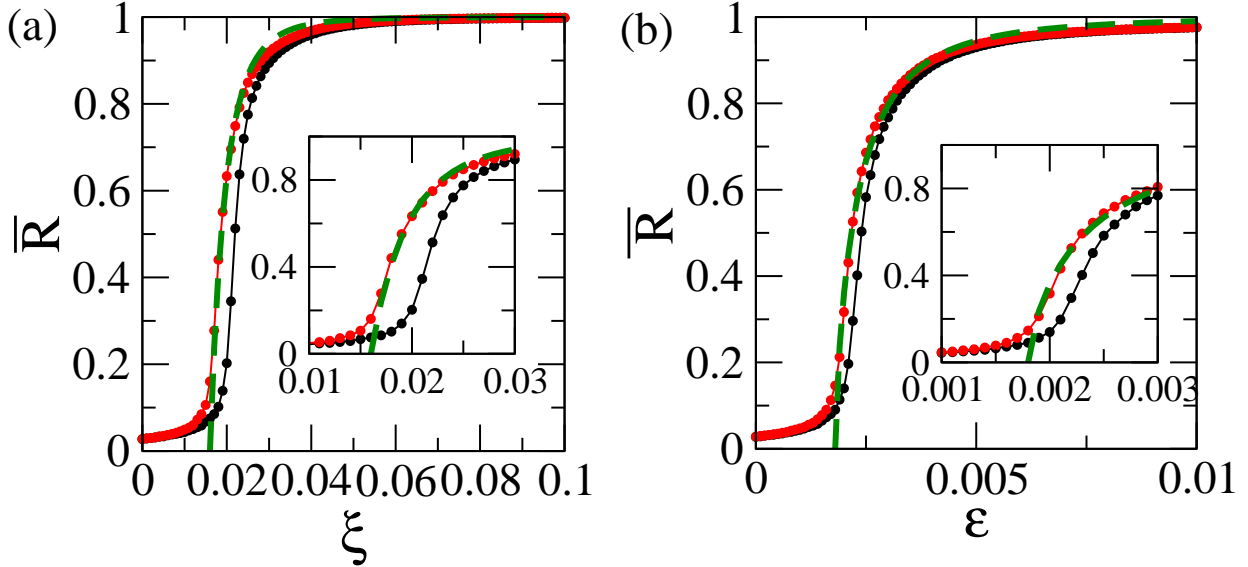}
\caption{\label{g} (color online) Variation of the order parameter magnitude with the coupling strength for a globally coupled (a), and a randomly coupled (b) network of $N = 1000$ Rulkov neurons and probability $p = 0.01$. Black and red lines stand, respectively, for a waterbag and a truncated Cauchy distribution of values of the $\alpha$ parameter in the $[4.1:4.3]$. The green curves represent a polynomial fit given by Eq. (\ref{Rfit}). The insets zoom the behavior near the transition to bursting synchronization.}
\end{figure}

The variation of the order parameter magnitude of the coupled network with the coupling strength above is depicted in Fig. \ref{g}(a), where we plot $\overline{R}$ as a function of $\xi$ when the values taken on by the parameter $\alpha^{(i)}$ are drawn from a waterbag PDF (black line) and a truncated Cauchy distribution (red line). Since we used randomly chosen initial conditions for the neurons $(x_0^{(i)},y_0^{(i)})$ we considered averages over $100$ different realizations of the initial pattern.

In both cases the behavior is qualitatively similar: for small coupling values there is no bursting synchronization and, after a critical coupling value $\xi_c \approx 0.020$ we begin to observe partial synchronization and, further on, complete phase synchronization. The approach to the latter differs with respect to the PDF ${\tilde g}(\alpha)$, being slightly faster for a truncated Cauchy distribution than for a waterbag PDF [see the inset of Fig. \ref{g}(a)].

The ``physical'' distance between two neurons does not play any role in globally coupled networks, since the contribution is counted in the same way for all nodes. This deficiency has been circumvented by considering a network in which the coupling strength decreases with the lattice distance as a power-law, with exponent $\varsigma$ \cite{outros1}. If $\varsigma \rightarrow 0$ the lattice distance does not matter and we recover the globally coupled case. As the exponent $\varsigma$ tends to infinity we approach a locally coupled network, where only the nearest-neighbor nodes have to be taken into account. 

Unlike global networks, random network present a typically small density of links. Random (Erd\"os-Renyi) networks are obtained by from $N$ initially uncoupled nodes and building $N_K$ links between randomly chosen pairs with a uniform probability $p$ (avoiding self-interactions) \cite{erdos}. Since the total possible number of links is $N(N-1)$ one has
\beq
\label{prober}
p = \frac{N_K}{N(N-1)}.
\eeq
The average path length of ER networks is typically very small since it scales as the logarithm of the network size ($L_{random} \sim \ln N$). Moreover the clustering coefficient of ER networks is likewise small since $C_{random} = 1/N$. 

Since the contribution of the coupling term is typically small we do not need here to rescale it as we did before. We show, in Fig. \ref{g}(b), the variation of the order parameter magnitude with the coupling strength $\varepsilon$, showing a qualitatively similar picture but with a considerably lower critical coupling strength $\varepsilon_c \approx 0.002$ for the transition from a non-synchronized to a partially synchronized bursting. If the probability $p$ is too small there may not occur bursting synchronization at all, and we require a minimum number of links to observe such phenomenon, as shown in Fig. \ref{g}(b), where $p = 0.01$ and we have $pN(N-1) = 9990$ links. 

\subsection{Small world networks}

Random networks of the Erd\"os-Renyi type represent one limit case of a spectrum of networks of which the other end comprises regular networks, which are lattices of $N$ nodes for which each node has links to its $z$ nearest and next-to-the-nearest neighbors, hence there are local connections only. The average path length of one-dimensional regular lattices is $L_{regular} \sim N$ and the clustering coefficient is given by $C_{regular} = [3(z-2)]/[4(z-1)]$ \cite{newman00}.

In between those limiting cases we have small-world networks, for which we typically have $L \gtrsim L_{random}$ and $C \gg C_{random}$. It is possible to obtain them from regular lattices essentially by two procedures: (i) Watts-Strogatz (WS) and (ii) Newman-Watts (NW) ones. WS networks are obtained from regular lattices by going through each of its links and, with some specified probability $p$, randomly rewiring that link, moving one of its ends to a new node randomly chosen from the rest of the lattice \cite{was}. 

The coordination number of the network is still $z$ on average, but the number of links per node may be greater or smaller than $z$ due to the possible existence of nonlocal shortcuts, in addition to the local links. One possible disadvantage of this construction is that, if $N$ is small enough, the rewiring process may create clusters disconnected to the rest of the network, for which nodes the path length would be obviously infinite. The latter problem can be circumvented in NW networks, that are constructed similarly to the WS ones, but the nonlocal shortcuts are added to the lattice with probability $p$, instead of being rewired \cite{nw}. 

A small-world network of either WS or NW types is essentially described by the probability of nonlocal shortcuts $p$. By computing both $L$ and $C$ we can get a range of $p$ for which the small-world property is fulfilled by the network. The small-world property requires the network to exhibit a relatively small path length while retaining an appreciable degree of clustering, i.e. the following conditions $L \gtrsim L_{random}$ and $C \gg C_{random}$ may hold for many different types of networks of the WS or NW types \cite{was}. The adjacency matrix for such networks is symmetric and band diagonal (with zero diagonal elements) and the non-diagonal parts are sparse, most of their elements being equal to zero. 

\begin{figure}
\includegraphics[width=0.9\textwidth,clip]{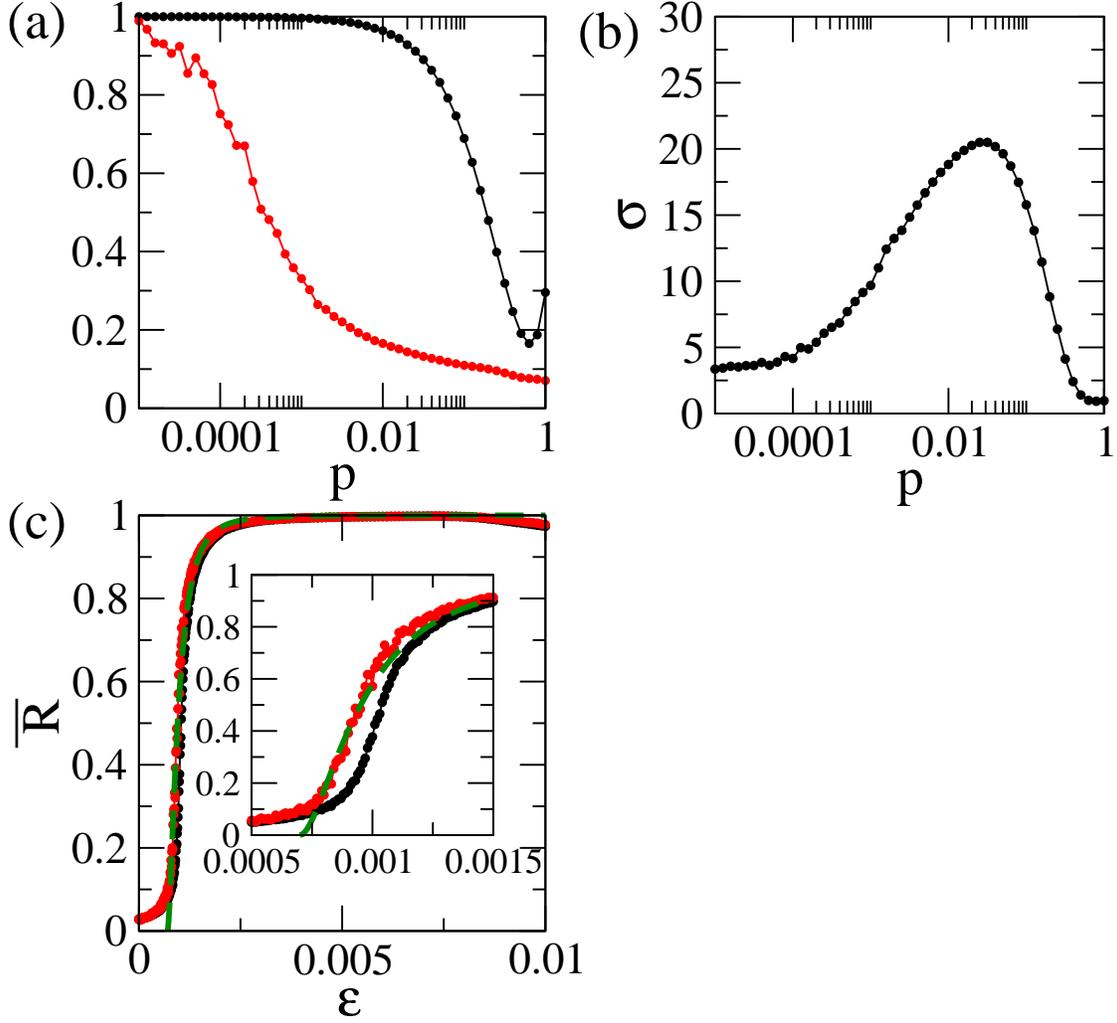}
\caption{\label{s} (color online) (a) Dependence of the ratios $C(p)/C(0)$ (black line) and $L(p)/L(0)$ (red line) with the probability of shortcuts in a small-world network of the Newman-Watts type with $N = 1000$ and $z = 20$; (b) Dependence of the merit figure $\sigma$ with $p$; (c) Variation of the order parameter magnitude with the coupling strength for a small-world network of Rulkov neurons with probability $p = 0.1$. Black and red lines stand, respectively, for a waterbag and a truncated Cauchy distribution of values of the $\alpha$ parameter in the $[4.1, 4.3]$. The green curve represents a polynomial fit given by Eq. (\ref{Rfit}). The insets zoom the behavior near the transition to bursting synchronization.}
\end{figure}

If the probability of shortcuts is zero we have the average path length and cluster coefficient of a regular networks, denoted as $L(0)$ and $C(0)$, respectively, for a network of $N=1000$ neurons with $z=20$ local connections. In Fig. \ref{s}(a) we plot the ratios $C(p)/C(0)$ and $L(p)/L(0)$ as a function of the probability of randomly chosen shortcuts $p$ in the NW procedure. The small-world property holds as long as the ratio $C(p)/C(0)$ is large and $L(p)/L(0)$ is small, what yields an interval $[0.01, 0.1]$. A further confirmation of this interval is provided by the merit figure $\sigma$, defined by 
\beq
\label{sigma}
\sigma=\frac{\kappa}{\lambda},
\eeq
where 
\beq
\label{ratios}
\lambda=\frac{L(p)}{L_{\mbox{\tiny rand}}}, \qquad \kappa=\frac{C(p)}{C_{\mbox{\tiny rand}}},
\eeq
in such a way that the larger is $\sigma$ the better the small-world property holds for the network. We plot the merit figure as a function of $p$ in Fig. \ref{s}(b), which indeed assumes larger values
in the interval $[0.01, 0.1]$. The variation of the order parameter magnitude for a small-world network with $p = 0.1$ is depicted in Fig. \ref{s}(c) as a function of $\varepsilon$, showing a transition to partial bursting synchronization for $\varepsilon_c \approx 0.001$. 

\subsection{Scale-free networks}

In scale-free networks the connectivity $k_i$ (the number of links per node $i$) satisfies a power-law PDF 
\beq
\label{sf}
P(k) \sim k^{-\gamma}, \qquad (\gamma > 1),
\eeq
in such a way that highly connected neurons are connected, on the average, with highly connected ones, a property also found in many social and computer networks \cite{barabasi,albert}. Functional magnetic resonance imaging experiments have suggested that some brain activities can be assigned to scale-free networks, with a scaling exponent $\gamma$ between $2.0$ and $2.2$, with a mean connectivity $<k> \approx 4$ \cite{chialvo01}. In fact, this scale-free property is consistent with the fact that the brain network increases its size by the addition of new neurons, and the latter attach preferentially to already well-connected neurons.

In this paper we use the Barab\'asi-Albert (BA) coupling prescription through a sequence of steps, starting from an initial random (Erd\"os-R\'enyi) network of size $N_0=23$ nodes and $23$ random connections \cite{albert}. Every step we add a new node to the network which makes two connections: the first is determined by a uniform probability of connection among the vertices in the network and the second link is chosen such that the probability of connection decays according with the degree of connectivity of each node. Hence this second link is more likely to be attached to the most connected node in the network than to the less connected one. When the network reaches the desired size $N$ we stop adding new nodes. 

\begin{figure}
\includegraphics[width=1.0\textwidth,clip]{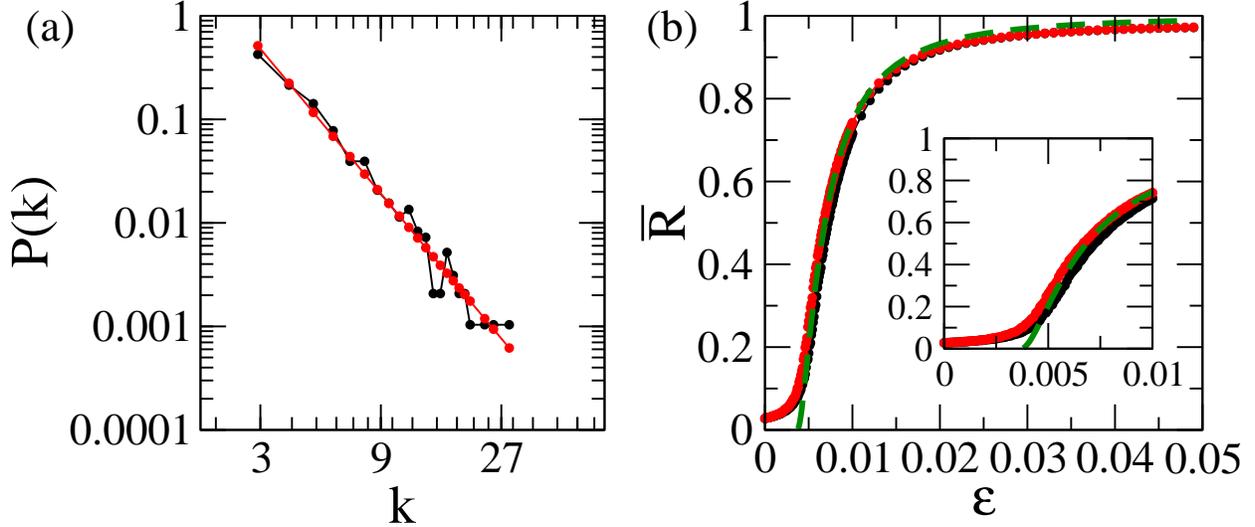}
\caption{\label{law} (color online) (a) Probability distribution function for the number of connections per site in a network obtained from BA procedure and $N = 1000$ nodes; (b) Variation of the order parameter magnitude with the coupling strength for a small-world network of Rulkov neurons with probability $p = 0.1$. Black and red lines stand, respectively, for a waterbag and a truncated Cauchy distribution of values of the $\alpha$ parameter in the $[4.1:4.3]$. The green curve represents a polynomial fit given by Eq. (\ref{Rfit}). The insets zoom the behavior near the transition to bursting synchronization.}
\end{figure}

We applied the BA procedure until the final network has $N = 1000$ nodes. We then obtained a numerical approximation for the PDF of connections per node $P(k)$, depicted in Fig. \ref{law}(a). The solid red line is a least squares fit giving a power-law dependence of the form (\ref{sf}) with an exponent $\gamma=2.9$, confirming that the network is scale-free indeed. The variation of the order parameter magnitude with $\varepsilon$ is plotted in Fig. \ref{law}(b), illustrating the transition to bursting synchronization with $\varepsilon_c = 0.004$. The behavior of the average order parameter magnitude with $\varepsilon$ after the transition to bursting synchronization at $\varepsilon_c$ can be fitted, when the frequency distribution is a truncated Cauchy PDF, by the following expression
\beq
\label{Rfit}
\overline{R} \approx {\left\lbrack 1- {\left(\frac{\varepsilon_c}{\varepsilon}\right)}^r \right\rbrack}^s,
\eeq
where the exponents are different according to the type of network [Table \ref{fitab}].

\begin{table}
\begin{tabular}{c|c|c|c|c}
\hline 
type & $\varepsilon_c$ & $r$ & $s$ & Obs.\\ 
\hline
global & $0.016/N$ & $4.5$ & $1.0$ & \\ 
\hline
Erd\"os-R\'enyi & $0.0017$ & $2.0$ &$1.0$ &  $K = 10000$ links \\ 
\hline
small-world & $0.00075$ & $4.0$ & $2.0$ & NW: $p = 0.1$, $z = 20$\\ 
\hline
scale-free & $0.004$ & $2.0$ & $0.7$ & BA: $N_0=23$, $\gamma = 2.9$\\
\hline
\end{tabular}
\caption{\label{fitab} Critical coupling strength and fitting parameters for some types of networks with $N=1000$ nodes and a truncated Cauchy distribution.}
\end{table}

\section{Generalized Kuramoto model}

We can summarize the results of the previous Section by stating that, if the coupling strength is large enough in a network of coupled Rulkov neurons, there will be a transition to bursting synchronization when $\varepsilon = \varepsilon_c$. The critical values of the coupling strength are different, however, depending on the type of network (keeping the network size constant, see Table \ref{fitab}). This leads to the question of how the type of network might determine the value of $\varepsilon_c$.

The answer to this question is elusive even for minimal models like the network of Rulkov maps considered in this work. However it can be formulated in a simpler model which has the virtue of being amenable to analytical methods like mean field theory, and which is the generalized Kuramoto model. Let us define a phase $\theta_i \in [0,2\pi)$ for the $i$th member of an assembly of $N$ oscillators connected by a network of which we know the adjacency matrix $A_{ij}$
\beq
\label{genkura}
\frac{d\theta_i}{dt} = \omega_i + \sigma \sum_{j=1}^{N} A_{ij} \sin(\theta_j-\theta_i), \qquad (i=1,\ldots,N)
\eeq
where $\sigma$ is the coupling strength and $\omega_i$ are natural frequencies which we randomly choose from a PDF $g(\omega)$ which we require to be unimodal and symmetric: $g(-\omega) = g(\omega)$. The generalized Kuramoto model (GKM) has been used to study oscillations in cortical circuits \cite{breakspear}, as well as properties as axonal delay and synaptic plasticity \cite{timms}. 

A number of recent works has considered analytically the onset of synchronization in the GKM on complex networks \cite{restrepo,arenasrev}. Two basic approximations have been made: (i) the adjacency matrix is symmetric ($A_{ji}=A_{ij}$); (ii) the number of connections per node is large enough ($k_i \gg 1$). In this case the critical coupling is given by
\beq
\label{sigmac1}
\sigma_{c1} = \frac{K_c}{\lambda_{max}},
\eeq
where $\lambda_{max}$ is the largest eigenvalue of the adjacency matrix and
\beq
\label{Kc}
K_c = \frac{2}{\pi g(0)}
\eeq
is the critical coupling strength of the classical Kuramoto model, which corresponds to the case of a globally coupled network \cite{kuramoto,strogatz,acebron}. Indeed, since $A_{ij} = 1$ for them it turns out that $\lambda_{max} = N-1$ and, at criticality, we have
\beq
\label{glokura}
\frac{d\theta_i}{dt} = \omega_i + \frac{K_c}{N-1} \sum_{j=1}^{N} A_{ij} \sin(\theta_j-\theta_i).
\eeq

A mean field analysis based on the same assumptions as above furnishes the following estimate for the critical coupling strength in the GKM \cite{arenasrev}
\beq
\label{sigmac2}
\sigma_{c2} = K_c \frac{<k>}{<k^2>},
\eeq
where $<k>$ and $<k^2>$ are the mean connectivity and variance, respectively, or the two first moments of the PDF $P(k)$. The limitations of this formula are clear, though, since it was derived by assuming that $k_i \gg 1$, which is hardly the case for small networks. An example of the possible inadequacy of this formula is the case of a scale-free network, for which the variance
\beq
\label{vark2}
<k^2> = \int_1^\infty dk \, P(k) \, k^2 \sim \int_1^\infty dk \, k^{2-\gamma},
\eeq
which diverges if $\gamma \le 3$, yielding an infinite value for the critical coupling strength. 

\section{Discussion}

In this Section we will compare results obtained from a network of coupled Rulkov networks and the GKM. Similarities between them have been previously observed but no direct connection between those models has been made so far \cite{jalili,ewandson,nordenfelt}. The starting point of our discussion is that the Kuramoto phase $\theta_i(t)$ can be identified with the bursting phase of a Rulkov neuron $\varphi_i(n)$. If the time discretization step $\tau$ is small enough it is immaterial if the model uses continuous time (like the GKM) or discrete time (like the Rulkov map). The similarity between these models can be proved on general grounds, when one is close to a globally phase synchronized state, the proof being sketched in an Appendix.

The bursting frequencies, which are time rates of the corresponding phases, should be similar to both Rulkov and Kuramoto networks. Hence we adjust the PDF's of the frequencies of the Rulkov model to comply with the symmetry requirements of the GKM. Let us first consider the waterbag distribution with $b = -a$:
\beq
\label{uniform1}
g_W(\omega) = \left \{ 
\begin{array}{ccc}
   \frac{1}{2a} & \mbox{for} & -a\leq \omega \leq a \\
   0 & \mbox{otherwise}
\end{array}
\right., 
\eeq
for which $g_W(0) = 1/2a$. The parameter $a$ for the GKM will be chosen so as to yield the same critical coupling strength which we numerically determined for the Rulkov network. For example, using the theoretical estimate (\ref{sigmac1}) we have at the critical point
\beq
\label{sigmac1a}
\sigma_{c1}^{(W)} = \frac{2}{\pi g_W(0) \lambda_{max}} = \frac{4a}{\pi \lambda_{max}},
\eeq
or, using (\ref{sigmac2})
\beq
\label{sigmac2a}
\sigma_{c2}^{(W)} = \frac{2}{\pi g_W(0)} \frac{<k>}{<k^2>} = \frac{4a}{\pi} \frac{<k>}{<k^2>}.
\eeq

\begin{table}
\begin{tabular}{c|c|c|c}
\hline 
network & $<k>$ & $<k^2>$ & $\lambda_{\mbox{\tiny max}}$ \\ 
\hline
global & $999$ & $999^2$ & $999$ \\ 
\hline
Erd\"os-R\'enyi & $10$ & $109.30$ & $11.019$ \\ 
\hline
small-world & $24.29$ & $594.83$ & $24.514$ \\ 
\hline
scale-free & $3.954$ & $25.058$ & $6.33$ \\
\hline   
\end{tabular}
\caption{\label{tabnets} Properties of the network models used in this paper.}
\end{table}

It is possible to repeat the arguments for the case of a truncated Cauchy distribution with $b = -a$, $\alpha_0 = 0$, and where we have set $\omega_0 = 0$ and $\gamma = a$ in such a way that we have 
\beq
\label{cauchy1}
g_T(\omega) = \frac{2}{\pi a} {\left\lbrack 1 + {\left(\frac{x}{a}\right)}^2 \right\rbrack}^{-1},
\eeq
with $g_T(0) = 2/(\pi a)$. This yields theoretical estimates for the critical point 
\bea
\label{sigmac1b}
\sigma_{c1}^{(T)} & = & \frac{2}{\pi g_T(0) \lambda_{max}} = \frac{a}{\lambda_{max}}, \\
\label{sigmac2b}
\sigma_{c2}^{(T)} & = & \frac{2}{\pi g_T(0)} \frac{<k>}{<k^2>} = a \frac{<k>}{<k^2>}.
\eea

Let us consider a network of $N = 1000$ Rulkov neurons with the following connection topologies: global, random (ER), small-world (NW), and scale-free (BA), whose parameters are listed in Table \ref{tabnets}. Using the numerically computed value of the critical coupling strength $\varepsilon_c$ in the place of the $\sigma_c$ in Eqs. (\ref{sigmac1a})-(\ref{sigmac2a}) and (\ref{sigmac1b})-(\ref{sigmac2b}) we can estimate the value of the parameter $a$ of the PDF of natural frequencies using both the waterbag and the Truncated Cauchy models, respectively [cf. Tab. \ref{tabval}]. In the case of a random (ER) network, for example, in which $\varepsilon_c = 0.020$ and $0.016$ for the waterbag and Cauchy PDFs, respectively, we found $a = 0.016$ and $0.017$. Similar results hold for other network topologies. The values of $\sigma_{c1}$ and $\sigma_{c2}$ present only slight differences. 

\begin{table}
\begin{tabular}{c|c|c}
\hline 
network & waterbag & truncated Cauchy \\ 
\hline
global & $0.016$ & $0.017$ \\ 
\hline
Erd\"os-R\'enyi & $0.017$ & $0.019$ \\ 
\hline
small-world & $0.017$ & $0.018$ \\ 
\hline
scale-free & $0.025$ & $0.025$ \\
\hline   
\end{tabular}
\caption{\label{tabval} Estimates for the value of parameter $a$ for two different PDFs and some types of networks.}
\end{table}

\begin{figure}
\includegraphics[width=1.0\textwidth,clip]{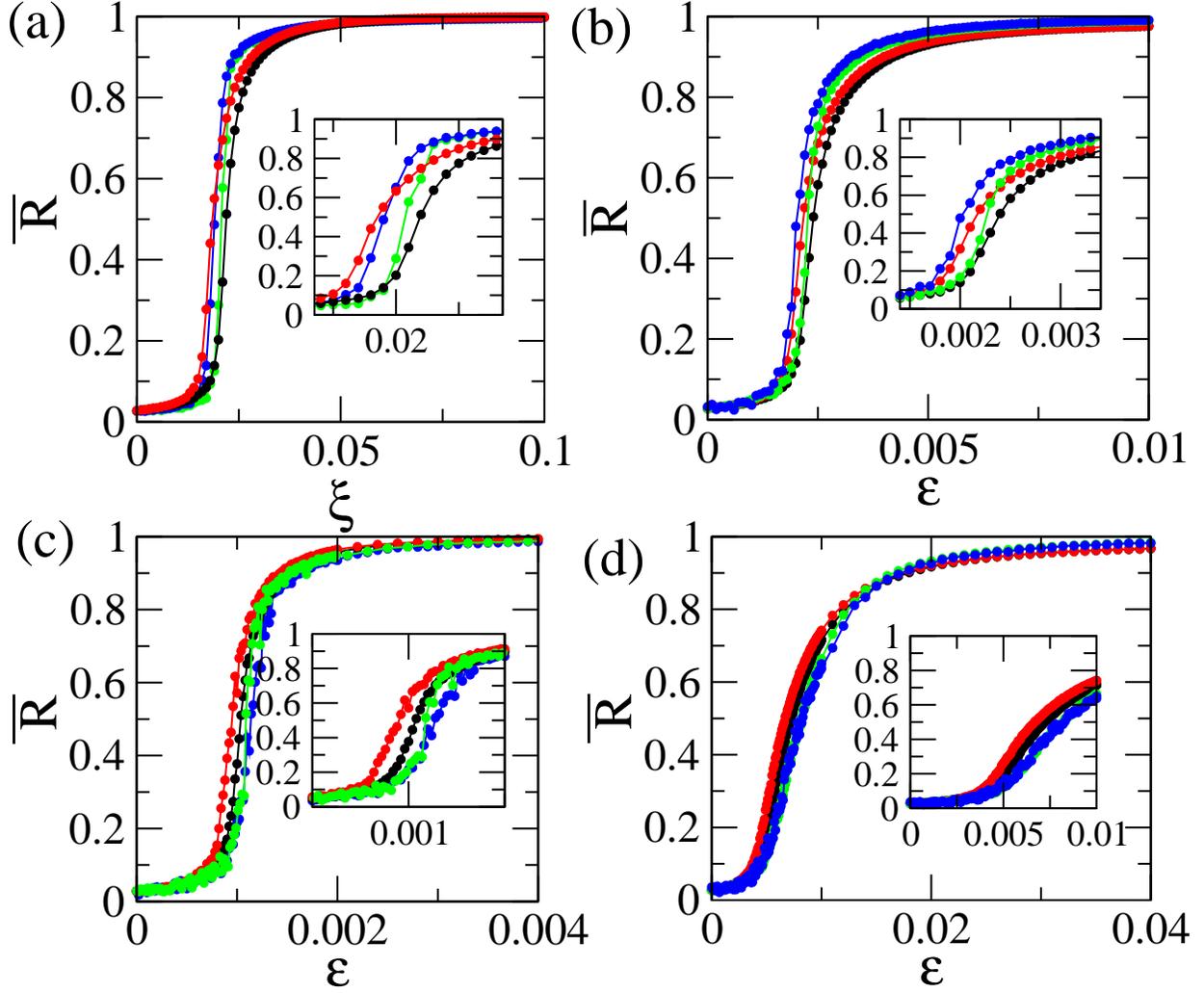}
\caption{\label{compfig} (color online) Order parameter magnitude vs. coupling strength for a network of Rulkov neurons (black: waterbag, red: Cauchy) and Kuramoto oscillators (green: waterbag, blue: Cauchy) with $N = 1000$ nodes, for (a) global coupling, (b) Erd\"os-R\'enyi; (c) small-world, and (d) scale-free.}
\end{figure}

After having chosen adequate values of the parameters appearing in the PDF of both Rulkov and Kuramoto models, we can compare the results for them with respect to the transition to bursting synchronization. In Figs. \ref{compfig}(a) to (d) we compute the order parameter magnitude for the Rulkov (black line: waterbag, red line: Cauchy) and Kuramoto (green line: waterbag, blue line: Cauchy) as a function of the coupling strength for the network topologies considered in this paper. The evolution of $R$ is nearly the same for both models, with small differences. For example the value of $R$ in the GKM reaches its  maximum slightly before (i.e. with smaller coupling strengths) the Rulkov network does. This precedence of GKM in comparison with Rulkov is observed for all connection topologies we have considered in this work. 

\section{Conclusions}

In this work we propose minimal models for the description of bursting neurons using some of the most used connection topologies for neuronal networks which can be used in numerical experiments. One of the simplest models that describe bursting is a two-dimensional discrete-time mapping proposed by Rulkov, and we considered networks of coupled Rulkov neurons using global, random (Erd\"os-Renyi), small-world (Newman-Watts), and scale-free (Barabasi-Albert) topologies. We can define a geometrical phase for coupled Rulkov neurons, such that bursting synchronization is equivalent to some form of phase synchronization. Using the time rate of the bursting phase we can define a bursting frequency and thus investigate frequency synchronization in such networks.

Using a complex order parameter we quantify the state of phase synchronization of the system. As the coupling strength is increased, there is a transition from a non-synchronized to a partially phase-synchronized state, which evolves to a completely phase-synchronized one for larger values of the coupling strength. We computed the corresponding critical values of the coupling strength for the connection topologies considered in this work.

The existence of a phase reduction for the coupled Rulkov neuron system suggests the usefulness of a Generalized Kuramoto model which, in the present context, can be considered a minimal model for bursting neurons. We presented in this paper some quantitative results that justify this claim. In order to compare both models, however, it is necessary to adjust the corresponding probability distribution functions for the natural frequencies. We did so by using theoretical estimates for the critical coupling strength for the Generalized Kuramoto model in different connection topologies. After choosing the PDF parameters in a suitable way the behavior of the order parameter is very similar for both the Rulkov and Kuramoto models. 

\appendix
\section{Phase reduction near global phase synchronization}

In this appendix we show that, in the scenario near a phase synchronized regime, one can perform a phase reduction from a network of coupled oscillators to a generalized Kuramoto model, what explains why the properties of a coupled Rulkov neuronal network are similar to the Kuramoto model. We stress that this similarity does not necessarily apply to the situation near a frequency synchronized regime, since the dynamical requirements are different. Our presentation follows closely that in Ref. \cite{ewandson} and the general treatment given in Ref. \cite{kuramoto}.

Let a general $D$-dimensional flow be given by
\beq
\label{flow}
\frac{d{\bf x}}{dt} = {\bf F}({\bf x}),
\eeq
\no where ${\bf x}$ is a $D$-dimensional vector in the system phase space and ${\bf F}$ a vector field. We assume that there is a stable period-$T$ orbit
\beq
\label{orbit}
{\bf x}_0(t) = {\bf x}_0(t+T).
\eeq
The ``slow'' dynamics along this periodic orbit can be described by a phase $\varphi({\bf x})$ which time evolution is given by
\beq
\label{phasedef}
\frac{d}{dt} \varphi({\bf x}) = \nabla_{\bf x} \varphi {\bf F}({\bf x}) = 1.
\eeq

Now let us consider a network of coupled oscillators with a slight mismatch in their parameters described by 
\beq
\label{coupled}
\frac{d}{dt} {\bf x}_i = {\bf F}({\bf x}_i) + {\bf f}_i({\bf x}_i) + \epsilon \sum_j a_{ij} {\bf V}({\bf x}_i,{\bf x}_j), \quad (i=1,2,\ldots N),
\eeq
\no where ${\bf f}_i$ is different for each oscillator and stands for the vector field part containing slightly mismatched parameters, $\epsilon$ is the coupling strength, $a_{ij}$ the adjacency matrix, and ${\bf V}({\bf x}_i,{\bf x}_j)$ is a coupling function. 

From (\ref{phasedef}), the slow phase $\varphi_i$ of the coupled oscillators is implicitely defined by the function
\beq
\label{Z}
{\bf Z}(\varphi_i) = \nabla_{\bf x} \varphi({\bf x}_0(\varphi_i)), 
\eeq
\no where ${\bf x}_0$ is a period-$T$ stable orbit. The time evolution of the phase is then governed by 
\beq
\frac{d}{dt} {\bf x} = 1 + \epsilon \sum_j a_{ij} {\bf Z}(\varphi_i) {\bf V}({\bf x}_i,{\bf x}_j) + {\bf Z}(\varphi_i){\bf f}_i({\bf x}_i),
\eeq
\no for $i=1,2,\ldots N$.

On introducing an auxiliary phase $\psi_i = \varphi_i - 1$ and using a time average over a period $T$, by keeping $\psi_i$ fixed we can write, in a first-order approximation, an equation governing the time evolution of $\psi_i$:
\beq
\label{psi}
\frac{d}{dt} {\psi}_i = {\tilde w}_i + \epsilon \sum_j a_{ij} \Gamma(\psi_i,\psi_j),
\eeq
\no where 
\bea
\label{w}
{\tilde w}_i & = & \frac{1}{T} \int_0^T {\bf Z}(t+\psi_i){\bf f}_i({\bf x}_i) dt, \\
\label{gama}
\Gamma(\psi_i,\psi_j) & = & \frac{1}{T} \int_0^T {\bf Z}(t+\psi_i){\bf V}({\bf x}_0(t+\psi_i),{\bf x}_0(t+\psi_j)) dt, 
\eea
\no play the roles of frequencies and coupling functions, respectively, for the auxiliary phases $\psi_i$. 

If we are close to a global phase synchronized state, for which $\varphi_i \approx \varphi_j$ for any pairs of oscillators $(i,j)$, we have an approximate form for the coupling function $\Gamma$. Introducing the time-dependent variable $\zeta = t + \psi_j$ and supposing that $\psi_j \ll T$ there results
\beq
\label{gama2}
\Gamma(\psi_i,\psi_j) = \frac{1}{T} \int_0^{T+\psi_j} {\bf Z}(\zeta+\psi_i-\psi_j) {\bf V}({\bf x}_0(\zeta+\psi_i-\psi_j),{\bf x}_0(\zeta+\psi_j)) dt.
\eeq

By expanding the coupling function ${\bf V}$ in a power series and assuming that ${\bf Z}$ is nearly constant over the periodic orbit ${\bf x}_0$, we get
\beq
\label{gama3}
\Gamma(\psi_i,\psi_j) \approx a + b (\psi_i-\psi_j),
\eeq
\no where 
\bea
\label{a}
a & = & \frac{1}{T} \int_0^T {\bf Z} {\bf V}({\bf x}_0(\zeta),{\bf x}_0(\zeta)) d\zeta, \\
\label{b}
b & = & \frac{1}{T} \int_0^T {\bf Z} {\left( \nabla_{\bf x}{\bf V}({\bf x},{\bf y}) \right)}_{{\bf x}={\bf y}={\bf x}_0(\zeta)} \frac{\partial {\bf x}_0}{\partial\zeta} d\zeta, 
\eea
\no Notice that this approximation holds whenever 
\beq
\label{cond}
\nabla_{\bf x}{\bf V}({\bf x},{\bf y}) \ne 0
\eeq
\no for any ${\bf x}$ and ${\bf y}$ belonging to the periodic orbit ${\bf x}_0$.

Substituting (\ref{gama3}) into (\ref{psi}) yields
\beq
\label{psi1}
\frac{d}{dt} {\psi}_i \approx {\tilde w}_i + s_i a + \epsilon b \sum_j a_{ij} (\psi_i-\psi_j),
\eeq
\no where $s_i = \sum_j a_{ij}$ is the intensity of the $i$th node of the network. As long as we deal with networks for which the intensities $s_i$ present only a small variation over the network, we can take $s_i$ as practically constant, i.e., independent of $i$. 

Moreover, since we are by hypothesis near a global phase synchronized state, the phase difference $\psi_i-\psi_j$ is small enough to justify the replacement $\psi_i-\psi_j \approx \sin(\psi_i-\psi_j)$, in such a way that the equation governing the time evolution of the auxiliary phases (near a phase-synchronized situation) is a generalized Kuramoto model
\beq
\label{kuramoto1}
{\dot{\psi}_i} \approx w_i + \varepsilon \sum_j a_{ij} \sin(\psi_i-\psi_j),
\eeq
\no where $w_i = {\tilde w}_i + a s_i$ and $\varepsilon=\epsilon b$.

\section*{acknowledgments}
This work was made possible through financial support from the Brazilian Research Agencies, CAPES and CNPq. Fabiano Ferrari thanks to Ruedi Stoop and the Institute of Neuroinformatics (UZH, ETHz) for having received him as a visiting student during 2014 through the program ``Science without boundaries''. We thank Dr. J. C. P. Coninck for fruitful discussions and helpful suggestions.

\end{document}